\documentclass[review]{elsarticle}

\usepackage{graphicx}   
\usepackage{subfigure}
\usepackage{color}
\usepackage[usenames,dvipsnames,svgnames,table]{xcolor}
\usepackage{amssymb,amsmath,amsfonts}
\usepackage{braket} 
\usepackage{upgreek}
\usepackage[colorlinks,linkcolor=cyan,anchorcolor=blue,citecolor=green]{hyperref}

\journal{Journal of \LaTeX\ Templates}

\begin{document}

\begin{frontmatter}
  
\title{A Holographic Static Transverse Polarization At Non-zero Temperatures}

\author{Lei Yin}
\ead{yinlei@m.scnu.edu.cn}
\author{Jialin Gao}
\address{Guangdong Provincial Key Laboratory of Nuclear Science, \\
  Institute of Quantum Matter, South China Normal University, Guangzhou 510006, China}

\begin{abstract}
  The analytical structure of a static transverse component of polarization tensor in complex momentum plane is numerically studied, which is holographically determined by a Einstein-Maxwell theory in asymptotically $D=3+1$ dimensional Anti-de Sitter spacetime. This strongly-coupled transverse polarization shows a pair of conjugate simple poles on the imaginary-axis at low temperature, which is different with the longitudinal component of the corresponding polarization and the counterpart in its weakly-coupled version.
\end{abstract}



\end{frontmatter}


\section{Introduction and Summary}\label{sec:introduction}

Fermi liquid theory has obtained a great achievement to explain a plenty of matters and phenomena in our real world and experiments. However, some remarkable exceptions, such as the normal phase of high-$T_c$ superconductors that behave as anomalous ``strange metal'', are beyond the predictions by it, those systems don't hold the entire characteristics of Landau's Fermi liquid, and the perturbative method becomes ineffective and unreliable, we call such systems non-Fermi liquid, which often involves strong-coupling interaction and shows weird transport properties or unusual low-temperature behaviors\cite{Howard2019}\cite{Liu2011}\cite{Faulkner}. In gauge/gravity duality, holographic dictionary supports a formulation to induce unperturbative correlations in diverse strongly-coupled models\cite{Witten1998}\cite{Maldacena1998}\cite{Aharony1999}, e.g. the retarded Green functions. By means of the calculations on the proper weakly interacting gravitational system, one can extract universal features of a lot of systems on quantum many body physics or high energy physics in strongly-coupled limit that are interpreted as a boundary theory of this gravitational system. In complex frequency plane, the existence of quasi-particles picture of excitation is connected to the distributions of singularities of two-point spinor operator. The non-Landau liquid phases are classified by scaling dimension of a spinor operator on boundary that determines the analytical structure on complex frequency plane, and its low-energy behavior is controlled by the IR fixed point \cite{Faulkner2011a}\cite{Sachdev2011}. It's shown that when the spinor operator is relevant in the IR CFT, there is no quasi-particle picture, while when this operator is irrelevant, the ordinary Landau liquid features survive in some special cases\cite{Cai2017b}. The quasi-particle description is attributed to the dynamics response of a system, therefore the transport properties, like conductivity, can be explored in this light. On the other hand, the static response indicates the decay and oscillatory behavior in spatial dimensions for a physical system. Friedel oscillation is a famous case, it shows a system with charged particles under the influence of a constant external electric field, the response of density distribution of those charged particles is described by the static density-density correlation in the view of quantum field theory.


Current-current correlation is another one of such typical cases, which is produced by the transverse component of polarization tensor. And the reaction of a material for a magnetic screening in an external field is derived by it via linear response theory. In weakly-coupled quantum field theory, the feature of magnetic screening is extracted by loop corrections via perturbative method. In this paper, we will consider the unperturbative static transverse component of polarization with a holographic strong-coupling at $T \ne 0$, which is dual to a Reinssner-Nordstr\"{o}m-AdS geometry. The holographically strongly-coupled system under consideration has finite chemical potential and beyond the probe limit. In weakly-coupled quantum field theory, the physical system will show a Friedel-type response, this analysis can be performed by the loop-diagram calculations and the investigation of analytical structure of transverse polarization in a complex momentum plane.

Interestingly, Ref.\cite{Anantua2013} studied two models of bosonic geometry, which turn out that the absence of the transverse Friedel-like oscillation in Einstein-Maxwell-dilaton theory without chemical potential and  D3/D5 system in probe limit, respectively. It seems that the two factors:  a complete back-reaction of Maxwell field on gravity and a finite density($\mu \ne0$) are indispensable to cause a Friedel-like oscillation triggered by a bosonic geometry.

With the consideration of the two factors, in our previous works \cite{Yin2016} and \cite{Yin2017}, we found that all singularities of the static polarization tensor are simple poles at $ T \ne 0$. Moreover, those asymptotic poles far away from the real momentum-axis form a Friedel-like pattern that has three characteristics:
  \begin{enumerate}
  \item the distribution is symmetric about the imaginary-axis; 
  \item  all singularities are situated on two straight-lines;
  \item any a pair of the nearest singularities on those lines shares the same spacing.
  \end{enumerate}
  
In the work \cite{Yin2019}, we further studied the explicit expressions of static transverse polarization on both small and large momentum expansion. It's proven that the strongly-coupled polarization is a meoromorphic function of momentum at $T \ne 0$; and at $T =0$, all simple poles condensate together and form a pair of branch-cuts, this pattern is very different with ones in weakly-coupled counterpart.

For the non-zero temperature situation, analytical work is hard to explore those region near real-axis. In this paper, we perform computational calculations to solve the solutions of the transverse polarization. In the Sec.\ref{sec:bulk-theory:-einst}, the setup for the bulk theory will be given under our consideration. In Sec.~\ref{sec:bound-cond-stat} the boundary conditions of gauge fields will be analyzed, which is used to compute  $\mathcal{C}_{yy}(\mathfrak{q})$. And in the last, Sec.~\ref{sec:discussion} we will discuss the meaning of our numerical results. Finally, two appendixes are given for some detailed of formulas.

\section{From Einstein-Maxwell Theory to Strongly-coupled Quantum Gauge Theory}
\label{sec:bulk-theory:-einst}

The Reinssner-Nordstr\"{o}m-AdS geometry is a solution of  Einstein-Maxwell theory with a negative cosmological constant $\Lambda$, the corresponding action reads:
\begin{align}
  S = \int \ \mathrm{d}^4 X \, \sqrt{-g}\; \bigg[  G_4\, \big( R - 2 \Lambda \big)-  K_4\, \big( F_{\mu \nu} F^{\mu \nu} \big)  \bigg] 
\label{eq:2}
\; ,\end{align}
which $R$ is the scalar curvature in $D=3+1$ dimensional AdS space-time with $\Lambda = - \frac{3}{L^2}$, $L$ is the AdS radius, it's easy to know the mass dimension of the coupling constant $G_4$ is $[G_4]=2$. $F_{\mu\nu}$ is the electromagnetic tensor,$F_{\mu\nu} = \partial_\mu A_\nu - \partial_\nu A_\mu$. We set the mass dimension of the gauge field $A_\mu$ is $[A_\mu]=1$, thus the mass dimension of  the coupling constant $K_4$ is dimensionless, $[K_4] = 0$. 

From Gubser–Klebanov–Polyakov–Witten (GKPW) formula, the one-particle irreducible correlation function,e.g. the polarization $\langle \mathcal{A}_\mu(x_1) \mathcal{A}_\nu(x_2) \rangle_\text{1PI}$, of a strongly-coupled system is given by
a functional derivative of classical action in bulk theory with respect to field $A_\mu$ with their boundary values $\mathring{A}_\mu$
\begin{align}
  \langle \mathcal{A}_\mu(x_1) \mathcal{A}_\nu(x_2) \rangle_\text{1PI}  = \frac{\delta^2 S_\mathrm{cl.}}{\delta \mathring{A}_\mu(x_1) \delta \mathring{A}_\nu(x_2)}    %
  \label{eq:1}
  \; ,\end{align}
where $\delta \mathring{A}_\mu = \lim\limits_{u \to 0} \delta A_\mu$ represents the variance of the AdS-boundary value of the gauge field, that is a solution to the bulk equation of motion of fluctuations for the gauge field. Due to the linear response theory, this two-point functions link to the density-density or current-current correlations,being dependent of their indexes. For our purpose, we concern the special component $\langle \mathcal{A}_y(x_1) \mathcal{A}_y(x_2) \rangle_\text{1PI}$, and denote its Fourier transform as $\mathcal{C}_{yy}$.

In order to consider the finite density, we find a on-shell solution  $(\bar g_{\mu\nu}, \bar A_\mu)$ with a nonzero value of $\lim\limits_{u \to 0}\bar A_0 = \mu$, that is interpreted as the chemical potential of strongly-coupled theory.

The on-shell solution of the holographic model \eqref{eq:2} for Einstein part is 
\begin{align}
  \mathrm{d}^2\bar s  = \bar g_{MN} \, \mathrm{d}x^M \ \mathrm{d}x^N = \left(\frac{L}{u}\right)^2 \left(-f(u) \ \mathrm{d}t^2 + \frac{1}{f(u)} \ \mathrm{d}u^2 +\ \mathrm{d}x^2 +\ \mathrm{d}y^2 \right)   \label{eq:9}
\; ,\end{align}
where  the Poincar\'{e} coordinates is used, $u=0$ is the AdS boundary and all spacetime variables $\{x^\mu;u\}$ are dimensionless that are rescaled from a dimensional coordinates $\{ X^\mu, r\}$ with $X^\mu = \frac{L^2}{r_+}\, x^\mu, r=\frac{r_+}{u}$, the metric function $f(u)$ can be derived: 
\begin{align}
  f(u)  = 1 - (1+Q^2) \, u^3 +  Q^2 \, u^4  \;,\label{eq:4}
\end{align}
which  $u=1$ is the outer horizon ; $Q= \frac{\mu L}{r_+}\sqrt{\frac{K_4}{G_4}}$ is dimensionless and related to the electric charge of the Reissner-Nordstr\"{o}m-AdS blackhole, $Q\in [0,\sqrt{3}]$.
Parameter $\mu$ is an integral constant, it presents in the on-shell solution for Maxwell part that the unique non-zero term is solved as
\begin{align}
  \bar A_t \ \mathrm{d}t = \mu\, L (1 - u) \ \mathrm{d}t  \label{eq:5}
\;, \end{align}
here $\mu, \; [\mu] =1$,  is the AdS boundary value of the gauge field $A_\mu$ and interpreted as the chemical potential of this system. By this on-shell solutions~\eqref{eq:4} and \eqref{eq:5}, we obtain the Hawking temperature 
\begin{align}
  T = \frac{\mu(3 - Q^2)}{4 \pi Q L} \sqrt{ \frac{K_4}{G_4} }  \;, \label{eq:13}
\end{align}
here $Q^2 \to 3$ corresponds to the zero temperature limit, and $\sqrt{\frac{K_4}{L^2G_4}}$ is a dimensionless coefficient, hence the Hawking temperature can be rescaled as if $\sqrt{\frac{K_4}{L^2G_4}}=1$ .

For the back-reaction between gauge field and metric field, we consider the fluctuations of the gauge vector field $a_\mu$ and metric tensor field $h_{\mu\nu}$ with the on-shell solution $\{\bar g_{\mu\nu}, \bar A_\mu \}$ as the background solution:
\begin{equation}
  \begin{aligned}
    g_{\mu \nu} &= \bar g_{\mu \nu} + h_{\mu \nu}  \;  ; \quad 
    A_\mu  &= \bar A_\mu + a_\mu
    \label{eq:3}
 \; , \end{aligned}
\end{equation}
where the radial gauge is taken to decrease gauge freedom degree, that is
\begin{align}
  h_{z \nu} = 0 , \qquad a_z = 0  , \quad \nu = \{t,x,y\}
  \; .\end{align}
We shall consider the solutions of the linearized Einstein-Maxwell equations as the first order fluctuation, more details shown in \cite{Edalati2010c} and \cite{Yin2019}, for our main purpose, the below coupled differential equations will be explored:
\begin{align}
  a_y'' + \frac{ f' }{ f } a_y' - \frac{ Q^2 \mathfrak{q}^2 }{ f } a_y - \frac{ \mu }{ f } h'^y_{\; t} &= 0 \label{eq:6}\\ 
  h''^y_{\;t}- \frac{ 2 }{ u } h'^y_{\; t} - \frac{ Q^2 \mathfrak{q}^2 }{ f } h_{\;t}^y - 4 \frac{ Q^2 }{ \mu } u^2 a_y' &= 0 \label{eq:7} \;. 
\end{align}
We will work in the static Fourier space, that is
\begin{align}
  h_{\mu \nu} (u; x ,y)  \; \sim \;  h_{\mu \nu} (u | q)  ; \; a_\mu(u ; x, y) \; \sim \; a_\mu(u | q) \; ,
\end{align}
and the dimensionless momentum $\mathfrak{q}= \frac{q}{\mu}$ is used, hence in the group of differential equations \eqref{eq:6} and \eqref{eq:7}, $a_y\equiv a_y(u|\mathfrak{q}) , h_{\;t}^y\equiv (u|\mathfrak{q})$ and the prime refers to a derivative with respect to $u$.

As a result, the variance of the AdS-boundary values of $A_\mu$ are obtained by solving the linearized Einstein-Maxwell equations. In view of linear response theory, the retarded Green's function has significantly physical meaning, which corresponds to the incoming condition near horizon in gauge/gravity duality. Alternatively, the outgoing horizon condition links to advanced Green's function on boundary theory\cite{Son2002a}. Our concerning static polarization should be taken as a limit $\omega \to 0$ for the retarded Green's function. In Fourier space, the static polarization reads
\begin{align}
  \mathcal{C}_{yy}(\mathfrak{q}) = 2\lim_{u \to 0} \sqrt{-g}g^{uu}g^{yy} \; \frac{a'_y(u|\mathfrak{q})}{a_y(u|\mathfrak{q})} \label{eq:8}
\; ,\end{align}
where $a'_y(u|\mathfrak{q})$ and $a_y(u|\mathfrak{q})$ are the solutions from Eqns. \eqref{eq:6} and \eqref{eq:7} with proper boundary conditions on AdS-boundary and horizon. The derivation of Eqn.~\eqref{eq:8} is put in ~\ref{sec:append-a:-deriv}.

\section{Boundary Conditions for Static Transverse Polarization}
\label{sec:bound-cond-stat}

Since the result of $\mathcal{C}_{yy}(\mathfrak{q})$ comes from the solution of Eqns.~\eqref{eq:6}\eqref{eq:7}, while the physical regular conditions are the permitted boundary conditions for those fields. In order to obtain this solutions, we need four independent boundary conditions.

An apparent boundary condition comes from the AdS-boundary condition of perturbative metric field $h_{\;t}^y \to 0$ by the requirement of GKPW formula \eqref{eq:1} that this second functional derivative operation produces polarization tensor. Due to the work of Ref.\cite{Yin2016}, we have an explicit AdS-boundary behavior:
\begin{align}
 h_{\; t}^y \sim O(u^3)  \;.   \label{eq:10}
\end{align}

Our system is a thermodynamic equilibrium, its temperature comes from Hawking radiation by RN-AdS black hole, to keep this topological structure of horizon for Hawking temperature \eqref{eq:13}, all perturbative metric fields should be vanishing on horizon, that means we have another two independent boundary conditions:
\begin{align}
  \tilde{h} \equiv \lim_{u \to 1} h_{\;t}^y &= 0 \; ; \label{eq:11} \\ 
  \tilde{h}' \equiv \lim_{u \to 1} {h_{\;t}^y}^\prime &= 0 \; . \label{eq:12}
\end{align}

In view of Gauge/Gravity Duality, the holographic polarization is originated from the fluctuations of gauge field on the AdS-boundary, $\mathring{a}_y(u|\mathfrak{q})$ and $\mathring{a}_y'(u|\mathfrak{q})$, this is different with performing variational principle to obtain classical equations of motion, that requires all variances, the first order fluctuations, of boundaries to vanish. The fluctuations $\mathring{a}_y(u|\mathfrak{q})$ and $\mathring{a}_y'(u|\mathfrak{q})$ are what we're interested in, that are the solutions of Eqns.~\eqref{eq:6} and \eqref{eq:7}, hence, the values of gauge field $a_y(u|\mathfrak{q})$ and $a_y'(u|\mathfrak{q})$ on horizon provide the boundary conditions for solving  Eqns.~\eqref{eq:6} and \eqref{eq:7}, while the regularity of the action \eqref{eq:2} determines those values. It's easy to know a special property in this $D=3+1$ dimensional bulk theory, due to Eqns.\eqref{eq:2} and \eqref{eq:9}, the corresponding effective action of gauge field reads
\begin{align}
  S_\text{EM eff.} = - 2 K_4 \int \ \mathrm{d}x^4 \big( \sqrt{-g} g^{uu}g^{yy} \big)\, F_{uy} \partial_u A_y = -2 K_4 \int  \ \mathrm{d}x^4 f(u) \ \partial a_y a_y \label{eq:16} \; ,
\end{align}
which shows those geometrical quantities from a $D=3+1$ dimensional metric field is reduced to a single metric function $f(u)$ that is null at horizon, $f(u)\big|_{u = 1} =0$, therefore, the physical regularity of the action only requires those fluctuations to keep non-infinite values on horizon:
\begin{align}
  \tilde{a}_y \equiv \lim_{u \to 1} a_y(u) = \text{Constant} \quad \text{or}  \quad \tilde{a}_y' \equiv \lim_{u \to 1} a_y'(u) = \text{Constant}  \;. \label{eq:14}
\end{align}

\begin{figure}[!htbp]
  \centering
  \subfigure[Use condition $\tilde{a}_y= 0$]{
    \begin{minipage}[!b]{0.45\linewidth}
      \includegraphics[width=2.2in]{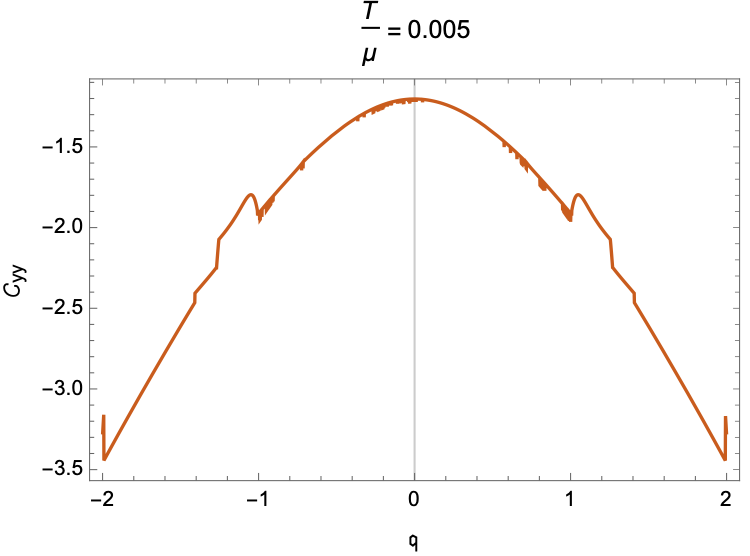}
    \end{minipage}%
  }
  \subfigure[Use condition $\tilde{a}'_y= 0$]{
    \begin{minipage}[!b]{0.45\linewidth}
      \includegraphics[width=2.2in]{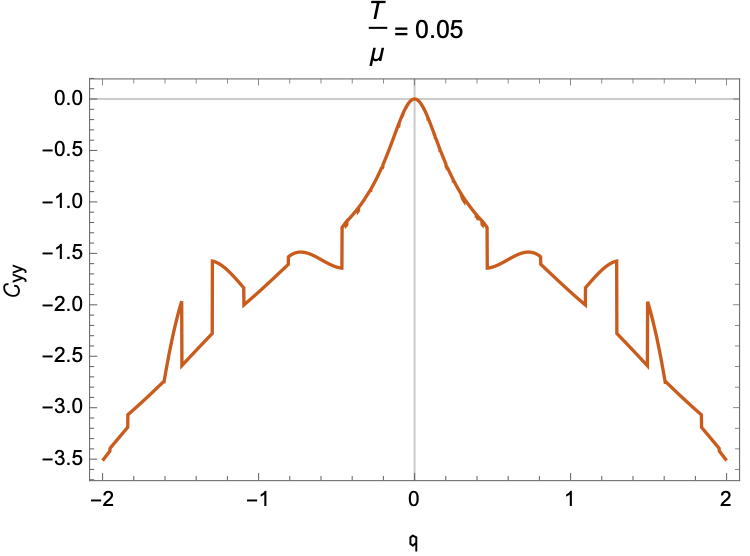}
    \end{minipage}%
  }
  \caption{The homogeneous conditions for the fluctuations of gauge field on horizon give rise to an unanalytical function $\mathcal{C}_{yy}$. }
  \label{fig: complexplot}
\end{figure}

From the analysis above, we have four independent boundary conditions for our numerical computation, to eliminate the doubt about the  validity of the conditions  \eqref{eq:10}\eqref{eq:11}\eqref{eq:12}\eqref{eq:14}, we firstly compute $\mathcal{C}_{yy}(\mathfrak{q})$ in real momentum $\mathfrak{q}$ expansion. Because of the analytical work in Ref.\cite{Yin2019}, we have a comparison to tell the validity of our computation. In small momentum expansion, it's proven that $\mathcal{C}_{yy}(\mathfrak{q})$ is an analytic function along the real-axis, and shows
\begin{align}
  \mathcal{C}_{yy}(\mathfrak{q}) \sim - \mathfrak{q}^2 + O(\mathfrak{q}^4) \;, \label{eq:15}
\end{align}
that means $\mathcal{C}_{yy}(\mathfrak{q})$ indicates a parabola opens down in the leading term of the small momentum expansion. However, those constant values on horizon in Eqn.~\eqref{eq:14} have two channels to distinguish themselves, if taking one of constants in Eqn~\eqref{eq:14} be zero as a boundary condition to solve Eqns. \eqref{eq:6} and \eqref{eq:7}, the computations show $\mathcal{C}_{yy}(\mathfrak{q})$ is not an analytic function with respect to $\mathfrak{q}$. We put two of our computation to demonstrate this point in Fig.\ref{fig: complexplot} .

Moreover, the two homogeneous conditions fail to connect to a small real-momentum expansion of $\mathcal{C}_{yy}(\mathfrak{q})$ in a manner of the parabola opening-down. As a result, the fluctuations on horizon should be finite values. Since Eqn.~\eqref{eq:6} and \eqref{eq:7} are homogeneous differential equations, this finite constant legitimately is rescaled to unit. Without loss of generality, we pick up
\begin{align}
 \tilde{a}_y = 1 \label{eq:17}
\end{align}
as one of four independent conditions for solutions. Based on the conditions \eqref{eq:10}\eqref{eq:11}\eqref{eq:12} and \eqref{eq:17}, in the Fig~\ref{fig:small-momentum} below, lots of $\mathcal{C}_{yy}(\mathfrak{q})$ on this leading term expansion are computed at various temperatures, which is in agreement with the analytical result:
\begin{figure}[!htb]
  \centering{\includegraphics[width=12.5cm]{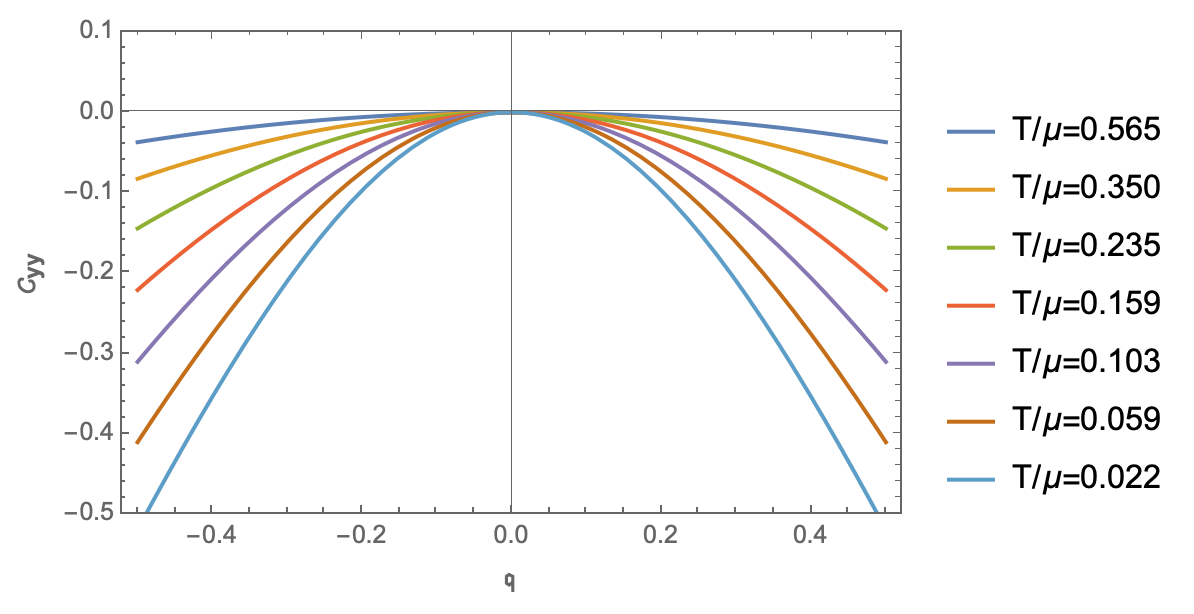}
  }
  \caption{ The graphs of the small momentum expansion $\mathcal{C}_{yy}(\mathfrak{q}) \sim - \mathfrak{q}^2 + O(\mathfrak{q}^4)$ in various temperatures manifest themselves as a sort of standard parabola with vertex at the origin.}
  \label{fig:small-momentum}
\end{figure}

It's worthwhile to note that since the holographic transverse polarization \eqref{eq:1} is represented as a ratio as shown in \eqref{eq:8}, therefore, in analytical work, for the small momentum expansion, we needn't to practically solve the equations \eqref{eq:6} and \eqref{eq:7}. We put a short review on this analytical result in ~\ref{sec:appendix-b:-review}. Further, the analytical result \eqref{eq:15} indicates $\mathfrak{q} =0$ is a zero of second order for $\mathcal{C}_{yy}(\mathfrak{q})$, we will confirm this feature in the below numerical computation.

Now, we are ready to perform the computation for $\mathcal{C}_{yy}(\mathfrak{q})$ with respect to a complex dimensionless momentum $\mathfrak{q}$. To investigate the structure of singularities, we will make the computation spread through a region in complex momentum plane. It's shown as:

\begin{figure}[!htbp]
  \centering
  \includegraphics[width=12.5cm]{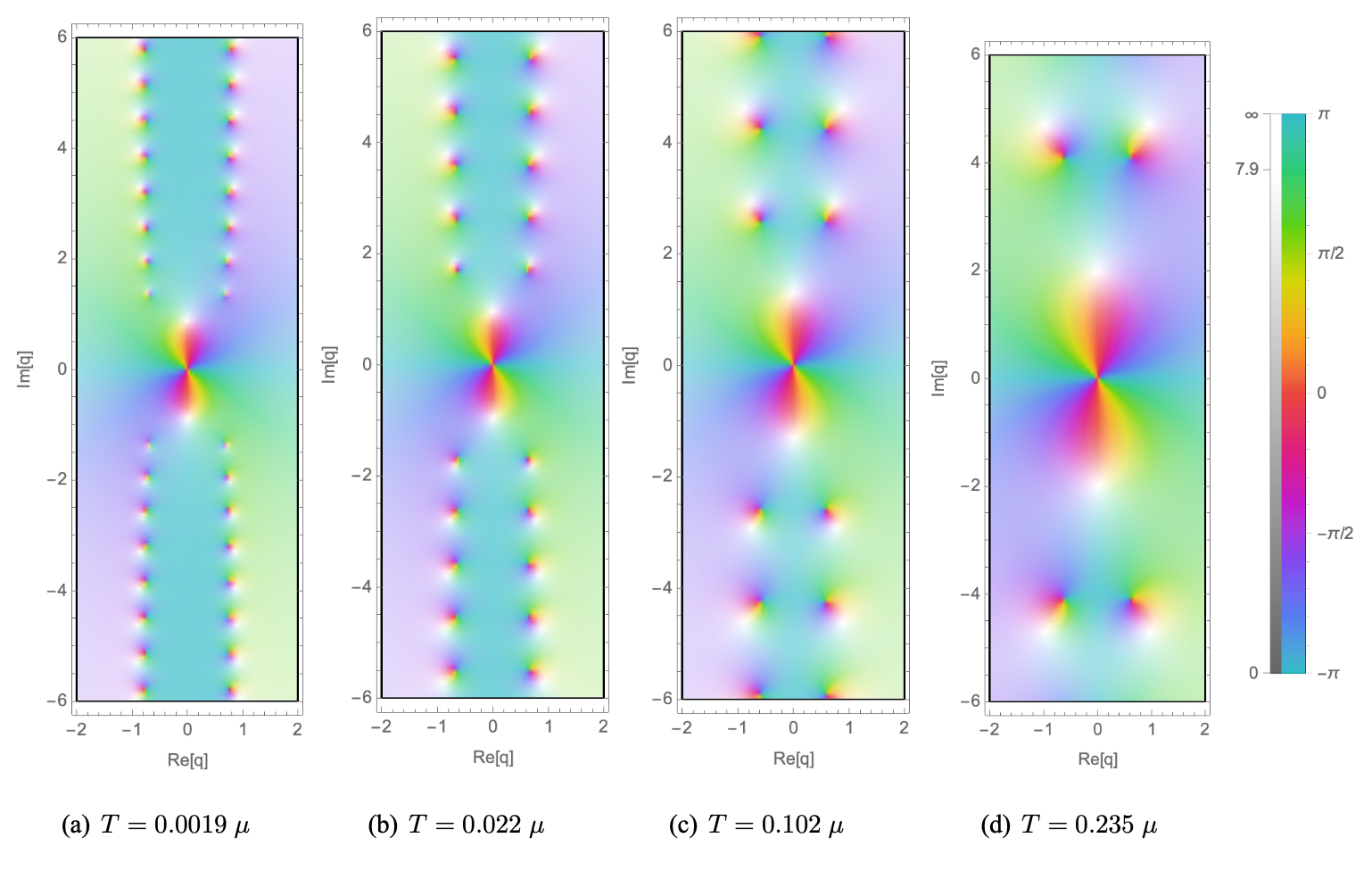}
  \caption{Static transverse polarization in various dimensionless temperature $\frac{T}{\mu}$. The color distribution refers to the argument of a complex function from $-\pi$ to $\pi$, shown on the right side. The color distribution around a point is counterclockwise, meaning that is a zero, alternatively, clockwise, meaning pole. If it is counterclockwise around a point twice, which means this point is a zero of second order. In above graphs, $\mathfrak{q}=0$ manifests itself a double  zero.}
   \label{fig: complexplot-1}
\end{figure}

\begin{figure}[!htbp]
  \centering
  \subfigure[$T= 0.438 \; \mu$]{
    \begin{minipage}[b]{0.45\textwidth}
      \includegraphics[width=2.5in]{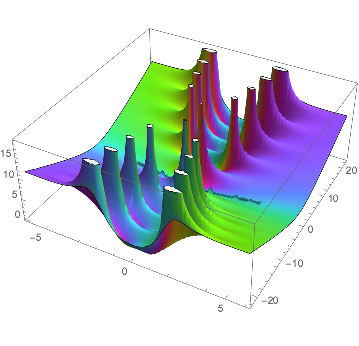}
    \end{minipage}%
  }
  \subfigure[$T=0.08 \; \mu$]{
    \begin{minipage}[b]{0.45\linewidth}
      \includegraphics[width=2.5in]{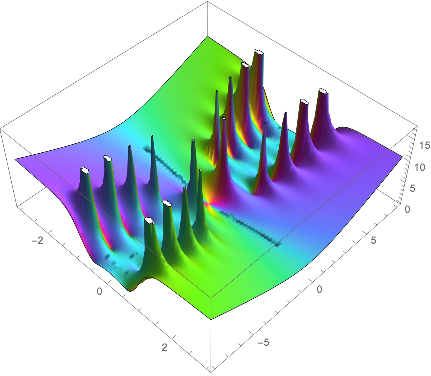}
    \end{minipage}%
  }

  \subfigure[$T=0.005 \; \mu$]{
    \begin{minipage}[b]{0.5\linewidth}
      \includegraphics[width=3.5in]{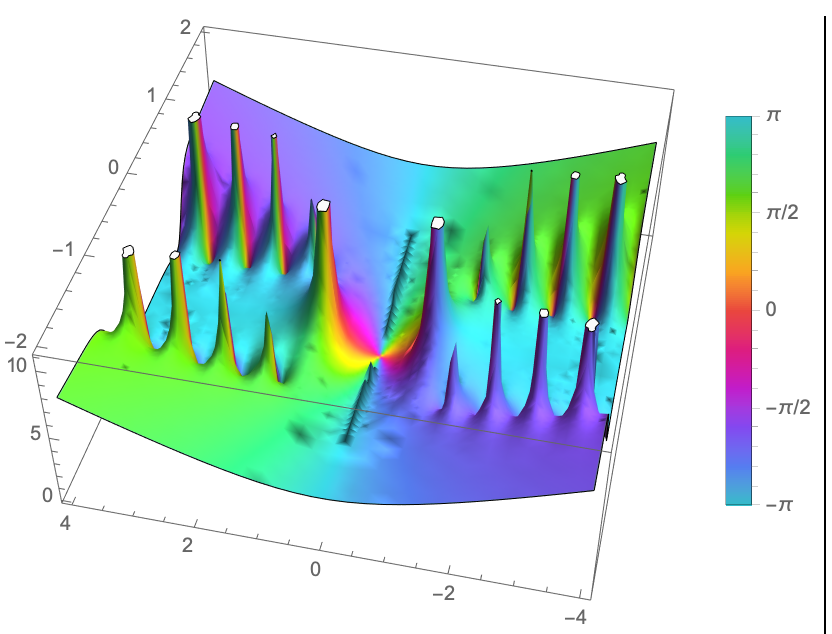}
    \end{minipage}%
  }
  \caption{The conjugate simple poles of static transverse polarization exist on the imaginary-axis as the temperature is lowered.}
  \label{fig: ComplexPlot3D}
\end{figure}

Besides a double zero at the origin has been shown in Fig.\ref{fig: complexplot-1}, we observe that the separation among poles becomes smaller as the temperature is lowering, this observation is in accordance with our previous analytical work mentioned in Sec.~\ref{sec:introduction}. Further, we find a sign of a pair conjugate simple poles on the imaginary axis. To make this message evidently, we perform some computations for 3D diagrams that clearly demonstrate the distribution of simple poles around the region of the real-axis even in very low temperatures, and are given in Fig.~\ref{fig: ComplexPlot3D}.

It's shown in Fig.~\ref{fig: ComplexPlot3D} that the presence of a pair of conjugate simple pole  on  the imaginary-axis at low temperatures, comparing to the chemical potential of this system, as well as the double zero at the origin.

\section{Discussion}
\label{sec:discussion}

By the numerical computations, we investigate the solutions of static transverse polarization in a Fourier space, which show a pair of conjugate simple poles on the imaginary-axis of complex momentum plane at a finite temperature. This consequence is very distinct from the longitudinal component of static polarization that is studied in Ref.\cite{Blake2015c}, that it shows all poles will move off the imaginary-axis below a critical temperature $T_c \approx 0.33 \, \mu$, and that no pole will present on the imaginary-axis gives rise to a Friedel oscillation for density-density correlation in a large length at $ T \ne 0$ . This is also quite different than the weakly-coupled counterpart that is no such poles situating on the imaginary-axis. It's worthwhile to explore more holographic models to check if there are a similar pair of poles on the imaginary-axis at non-zero temperature, and the distinction between longitudinal and transverse component of polarization. When turning it into the coordinate space by inverse Fourier transform, the pair of poles on the imaginary-axis will give a strong influence on a exponential decay behavior. A further research to explore in what manner is the exponential decay enforced is as well expected in future.

On the other hand, in the zero temperature limit, we have proven that the pattern of simple poles becomes branch-cuts symmetrically distributed on the complex plane, it's interesting to answer if there is no other singularity structure on the whole complex momentum plane except those branch-cuts. If the situation happened in such way, that means the absence of the conjugate imaginary poles at $T=0$. We are working on this problem in an analytical method, and the generalization of the investigation to more holographic models is expected.

\section*{Acknowledgement}
  We appreciate Prof.Ren Hai-cang for his helpful discussion.

\appendix
\section{Derivation to Transverse Polarization}
\label{sec:append-a:-deriv}

Starting with the Einstein-Maxwell theory in $D=d+1$ dimensional spacetime, its action is of the form in \eqref{eq:2}:
\begin{align}
  S_{\text{EM}} &= - 2 K_D \int \ \mathrm{d}^D X \sqrt{-g} \; F^{MN} \partial_M A_N = - 2 K_D \int \ \mathrm{d}u\, \mathrm{d}^d X \sqrt{-g} g^{uu} g^{yy} \; \partial_u a_y   \partial_u a_y  \\
                &=  \lim_{u \to 0} 2 K_D \int \  \mathrm{d}u\,\mathrm{d}X^d \sqrt{-g} g^{uu}g^{yy}\; \partial_u  \mathring{a_y} \mathring{a_y} \; + \boxed{\text{Horizon Part + E.O.M}}\;, \label{eq:18}
\end{align}
where an integration by part has been performed, noting $u=0$ be the AdS-boundary and for the polarization on surface theory, the effective part is only the first term in Eqn.~\eqref{eq:18}. Here, the metric tensor is of the form in \eqref{eq:9}, but to be generalized to a higher dimensions, also, the radial gauge $a_u \equiv 0$ is used. For some simple situation, we can take a coordinate to make the direction of momentum along the $y$-axis, $|\vec{p}| = p_y=q$  , that will lead to a simpler Fourier transform in $d$-dimensional surface theory:
\begin{align}
  a_\mu(u;x^\mu) = \int \frac{\mathrm{d}^d p}{(2\pi)^d} e^{\mathrm{i} p\cdot x} \; a_\mu(u | p) \; , \label{eq:22}
\end{align}
putting into Eqn.~\eqref{eq:18}, after some calculations, the Action is written in Fourier space:
\begin{align}
  S_{\text{EM}} \sim 2 K_D \lim_{u \to 0}  \sqrt{-g}g^{uu}g^{yy} \, \int \left( \frac{\mathrm{d} p}{2 \pi} \right)^d a'_y(u | p) a_y(u| -p) \;. \label{eq:19}
\end{align}

The last term in Eqn.~\eqref{eq:19} has a simple variant in static case:
\begin{align}
  a(u| -p) = a^\star(u|p) = A(u|q) \;. 
\end{align}
The retarded Green's function is a complex function in general case, but when in the static case, it becomes a real function, and due to our special choice of frame, the spatial momentum is expressed $q$. By means of the above relation, putting back to Eqn.~\eqref{eq:19}, in $D=3+1$ dimensions, we have
\begin{align}
  S_{\text{EM}} \sim 2K_4 \lim_{u \to 0} \sqrt{-g}g^{uu}g^{yy} \; \int \frac{\mathrm{d}p}{2\pi} \, \frac{\mathring{a}_y'(u|q)}{\mathring{a}_y(u|q)} \, \mathring{a}_y(u|q)\mathring{a}_y(u|q)  ; , \label{eq:20} 
 \end{align}
 this expression is suit for the consideration of linearized fluctuations. After 2nd functional derivative, the static transverse polarization is obtain
 \begin{align}
   \mathcal{C}_{yy} &= 2 K_4\lim_{u \to 0} \sqrt{-g}g^{uu}g^{yy} \; \frac{\mathring{a}_y'(u|q)}{\mathring{a}_y(u|q)} \,\\
                    &= 2K_4 \frac{\mathring{a}_y'(u|q)}{\mathring{a}_y(u|q)} \;, \label{eq:21}
 \end{align}
 where the spatial momentum $q$ must be rescaled to a dimensionless momentum in the Fourier transform \eqref{eq:22}, since here the coordinate variable is dimensionless, $q$ has the same meaning as $\mathfrak{q}$ we was using in this paper if it's rescaled by $\mathfrak{q}=\frac{q}{\mu}$. Finally, the result \eqref{eq:21} is equal to Eqn. \eqref{eq:8}.

\section{Review on the analytical result on small momentum expansion of $\mathcal{C}_{yy}(\mathfrak{q})$}
\label{sec:appendix-b:-review}

The small momentum expansion for $\mathcal{C}_{yy}(\mathfrak{q})$ is studies in Ref.\cite{Yin2019}, the key steps is to find the explicit expressions of $\mathring{a}_y(\mathfrak{q})$ and  $\mathring{a}_y'(\mathfrak{q})$ by not solving the Eqns.~\eqref{eq:6}\eqref{eq:7} directly, but via expanding the master-field equation in term of the momentum in order to obtain a pair of inhomogeneous equations, see Eqn.(25) Ref.\cite{Yin2019}, by means of the variation of parameters method, one can get
\begin{align}
  \mathring{a_y} &= \frac{1}{ 2 Q^2 Z} \frac{1}{k}   \bigg[  a_-^{(1)} \mathfrak{q}^2 - a_+^{(1)} 4 Z^2 \bigg]  \\
\mathring{a_y}' &= \frac{1}{ 2 Q^2 Z^2} \frac{1}{k}   \bigg[  a_-^{(1)} 4 Z^2  -  a_+^{(1)} \big( -4 Z^2 - (Z^2(Q^2 +3) -1) \mathfrak{q}^2 \big) \bigg] \; , \label{eq:23}
\end{align}
where $Z= \frac{3}{4}(1+Q^{-2})$, and which $k$ is related to the momentum that will be cancelled in the expression of $\mathcal{C}_{yy}(\mathfrak{q})$. $a_\pm$ are two integral constants, which is found $a_- = -a_+$ in Eqn.(69) of Ref.\cite{Yin2019}. The result \eqref{eq:23} can be straightly obtained by Eqn. (71) in Ref.\cite{Yin2019}. Please note that there is a typo in the formula (71) in Ref.\cite{Yin2019}, comparing to \eqref{eq:23}.

According to the result \eqref{eq:21}, we have
\begin{align}
  \mathcal{C}_{yy}(\mathfrak{q}) \sim - 2 K_4 \frac{Z^2(Q^2+3)-1}{Z^3} \mathfrak{q}^2 + O(\mathfrak{q^4}), \label{eq:24}
\end{align}
which shows that the graph of the small momentum expansion is a parabola opens down.

\providecommand{\href}[2]{#2}\begingroup\raggedright\endgroup

\end{document}